# The interstellar chemistry of $H_2C_3O$ isomers.

*Jean-Christophe Loison[1,2]\*, Marcelino Agúndez[3], Núria Marcelino[4], Valentine Wakelam[5,6], Kevin M. Hickson[1,2], José Cernicharo[3], Maryvonne Gerin[7], Evelyne Roueff[8], Michel Guélin[9]*

\*Corresponding author: jean-christophe.loison@u-bordeaux.fr

[1] *Univ. Bordeaux, ISM, UMR 5255, F-33400 Talence, France*
[2] *CNRS, ISM, UMR 5255, F-33400 Talence, France*
[3] *Instituto de Ciencia de Materiales de Madrid, CSIC, C\ Sor Juana Inés de la Cruz 3, 28049 Cantoblanco, Spain*
[4] *INAF, Osservatorio di Radioastronomia, via P. Gobetti 101, 40129 Bologna, Italy*
[5] *Univ. Bordeaux, LAB, UMR 5804, F-33270, Floirac, France.*
[6] *CNRS, LAB, UMR 5804, F-33270, Floirac, France*
[7] *LERMA, Observatoire de Paris, École Normale Supérieure, PSL Research University, CNRS, UMR8112, F-75014, Paris, France*
[8] *LERMA, Observatoire de Paris, PSL Research University, CNRS, UMR8112, Place Jules Janssen, 92190 Meudon, France*
[9] *Institut de Radioastronomie Millimétrique, 300 rue de la Piscine, 38046, St. Martin d'Heres, France*

We present the detection of two $H_2C_3O$ isomers, propynal and cyclopropenone, toward various starless cores and molecular clouds, together with upper limits for the third isomer propadienone. We review the processes controlling the abundances of $H_2C_3O$ isomers in interstellar media showing that the reactions involved are gas-phase ones. We show that the abundances of these species are controlled by kinetic rather than thermodynamic effects.

1 Introduction

The formation of Complex Organic Molecules (COMs) in interstellar media, particularly in dense molecular clouds, is a challenging issue. Until recently, the paradigm has been that COMs were formed on grains through surface chemistry (Tielens & Hagen 1982) and then released into the gas-phase in warm environments through thermal desorption (Herbst & Van Dishoeck 2009). However this mechanism is incompatible with cold environments where COMs have been detected (Bacmann *et al.* 2012, Cernicharo *et al.* 2012, Vastel *et al.* 2014, Agúndez *et al.* 2015). In addition, the mobility of heavy radicals on grain surfaces is low at typical dark cloud temperatures. Vasyunin & Herbst (2013) hypothesized that these species could be formed in the gas-phase through radiative association reactions between precursors initially formed on ices before being released into the gas-phase by the reactive desorption mechanism. This mechanism is based on the principle that the exothermicity of surface chemical reactions promotes the desorption of the product species (Garrod *et al.* 2007). In their model, the $CH_3OCH_3$ molecule, for instance, is formed by the reaction $CH_3O + CH_3 \rightarrow CH_3OCH_3 + h\nu$. Ruaud *et al.* (2015) showed that the formation of COMs could be very efficient at the surface of the grains if new processes are considered such as low temperature Eley-Rideal mechanisms and the direct reaction of physisorbed species with the substrate. Even considering a low efficiency for reactive desorption, Ruaud *et al.* (2015) were able to reproduce the observed abundances of COMs in cold dark clouds. Balucani *et al.* (2015) proposed an original way to produce $CH_3OCH_3$ and $HCOOCH_3$ requiring large elemental abundances of chlorine and fluorine, similar to the ones observed in diffuse media. However fluorine and chlorine atoms may be partly depleted in dense molecular clouds (Blake *et al.* 1986, Kama *et al.* 2015), limiting the effectiveness of this mechanism (Wakelam & Herbst 2008). Acharyya *et al.* (2015) showed that the low temperature reaction between OH and methanol induced by tunneling (Shannon *et al.* 2013) may be an efficient way to produce methoxy, a molecule detected in cold dark clouds (Cernicharo et al. 2012), and then possibly some COMs, in cold and dense interstellar clouds.

We present here the detection of two $H_2C_3O$ isomers, propynal (HC≡C-CH=O) and cyclopropenone (c-$C_3H_2O$) alongside viable chemical routes for their formation. The third isomer propadienone has been searched as well but not detected. In Table 1 we show the three stable $H_2C_3O$ isomers with some of their characteristics (we specifically note the particular geometry of propadienone).

Table 1: $H_2C_3O$ isomers characteristics

| | Relative Energy (kJ/mol) (Loomis *et al.* 2015) | µ (Debye) |
|---|---|---|
| propadienone: 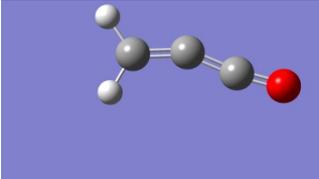 | 0 | $\mu_a$ = 2.156 ± 0.003<br>$\mu_b$ = 0.7914 ± 0.0006<br><br>(Brown *et al.* 1981) |
| propynal: | +22.7 | $\mu_a$ = 2.359 ± 0.018<br>$\mu_b$ = 1.468 ± 0.022 |

| | | | (Brown & Godfrey 1984) |
|---|---|---|---|
| cyclopropenone: 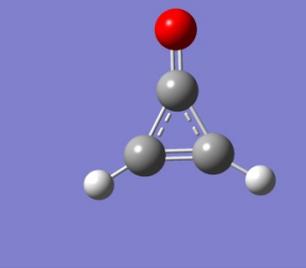 | | +45.0 | 4.39 ± 0.06 (Benson *et al.* 1973) |

The H$_2$C$_3$O family is interesting because even if propadienone is the most stable isomer (Karton & Talbi 2014, Loomis et al. 2015), only propynal and cyclopropenone have been detected in molecular clouds. Propynal has been previously observed in TMC-1 at cm-wavelengths (Irvine et al. 1988, Ohishi & Kaifu 1998) as well as in Sagittarius B2 (Turner 1991, Hollis *et al.* 2004, Loomis et al. 2015). The rotational spectroscopy of this asymmetric top molecule is well known from laboratory experiments (Barros *et al.* 2015). Cyclopropenone, the cyclic isomer, was first observed in Sagittarius B2 by Hollis et al. (2006) and was later on found in the cold dense molecular cloud L1527 (Tokudome *et al.* 2013). The rotational spectrum of cyclopropenone has been characterized in the laboratory (Guillemin *et al.* 1990). Propadienone has a peculiar rotational spectroscopy in which the bent skeleton and the small barrier to linearity split levels into two tunneling states (Brown *et al.* 1987); see also the Cologne Database for Molecular Spectroscopy[1] (Müller *et al.* 2005), resulting in a densely populated rotational spectrum, which makes its detection difficult. Previous searches for this isomer in various molecular sources (Irvine et al. 1988, Brown *et al.* 1992, Loomis et al. 2015) have been unsuccessful.

The astronomical observations are presented in Section 2. The chemical network to describe the chemistry of H$_2$C$_3$O isomers is based on the new version of kida.uva (Wakelam *et al.* 2015) with a thorough review of the various reactions producing and consuming neutral and ionic C$_3$H$_{x=0-8}$ in the gas-phase and on grains. This review will be presented in a forthcoming paper and we describe here only the reactions controlling the formation and destruction of H$_2$C$_3$O isomers (in Section 3). Our conclusions are presented in Section 4.

---

[1] http://www.astro.uni-koeln.de/cdms

2 Observations

Observations toward seven cold dark clouds were carried out with the IRAM 30m telescope using the frequency-switching technique and the EMIR 3 mm receiver connected to a fast Fourier transform spectrometer providing a spectral resolution of 50 kHz. The observations of TMC-1 and B1-b are part of a 3 mm line survey (Marcelino *et al.* 2009), a good part of which was observed between January and May 2012 (see details in (Cernicharo et al. 2012). In the cases of L483, Lupus-1A, L1495B, L1521F, and Serpens South 1a, the observations were carried out from September to November 2014 in selected frequency ranges across the 3 mm band (Agúndez et al. 2015). Examples of the observed spectra are shown in Fig. 1.

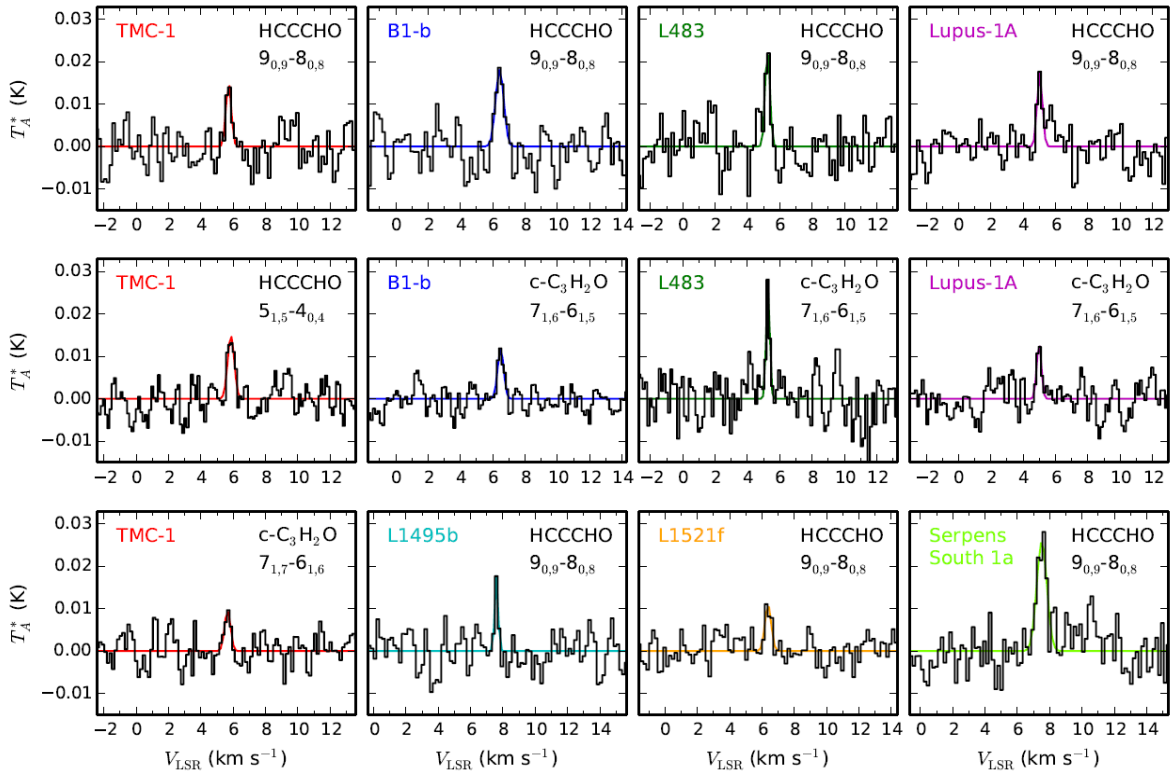

Figure1: Observed line profiles of c-$C_3H_2O$ and HCCCHO. The observed positions are: TMC-1 $\alpha_{2000.0}$ = $04^h41^m41^s.9$ $\delta_{2000.0}$ = +25°41'27".0; B1-b $\alpha_{2000.0}$ = $03^h33^m20^s.8$ $\delta_{2000.0}$ = +31°07'34".0; L483 $\alpha_{2000.0}$ = $18^h17^m29^s.8$ $\delta_{2000.0}$ = –04°39'38".3; Lupus-1A $\alpha_{2000.0}$ = $15^h42^m52^s.4$ $\delta_{2000.0}$ = –34°07'53".5; L1495B $\alpha_{2000.0}$ = $04^h15^m41^s.8$ $\delta_{2000.0}$ = +28°47'46".0; L1521F $\alpha_{2000.0}$ = $04^h28^m39^s.8$ $\delta_{2000.0}$ = +26°51'35".0; Serpens South 1a $\alpha_{2000.0}$ = $18^h29^m57^s.9$ $\delta_{2000.0}$ = –01°56'19".0.

TMC-1 was observed at the position of the cyanopolyyne peak (CP). The $H_2$ volume density has been estimated equal to $8 \times 10^4$ $cm^{-3}$ by Pratap et al. (1997), but in the range of 1-4 × $10^4$ $cm^{-3}$ by Snell et al. (1982) and close to $3 \times 10^4$ $cm^{-3}$ by Lique et al. (2006). The kinetic temperature is 10 K, according to Pratap et al. (1997) and 8 K according to Lique et al. (2006). The column density of $H_2$ molecules in TMC-1 CP is $10^{22}$ cm$^{-2}$ (Cernicharo & Guélin 1987). The Barnard 1 (B1) cloud was observed at a position between the two continuum sources B1-bN and B1-bS, where $N(H_2)$ is $1.3 \times 10^{23}$ cm$^{-2}$ (Hirano *et al.* 1999). The kinetic temperature and volume density in the high density clumps of B1 are 12 K and $(3.7-7) \times 10^4$ cm$^{-3}$ (Bachiller *et al.* 1990). Note however that Daniel et al. (2013) derive a steep density profile for B1, with values around $10^5$ cm$^{-3}$ at scales of ~10 arcsec. L483 is a cold dense cloud around a Class 0 source, with a column density of $H_2$ of $3 \times 10^{22}$ cm$^{-2}$ (Tafalla *et al.* 2000). The volume density and kinetic temperature derived by Jorgensen et al. (2002) are $3.4 \times 10^4$ cm$^{-3}$ and 10 K. The starless core Lupus-1A was discovered recently by Sakai et al. (2010) to be a rich source of carbon chain molecules. The kinetic temperature in the core is 14 K and the column density of $H_2$ is $1.5 \times 10^{22}$ cm$^{-2}$ (Agúndez et al. 2015). The volume density has not been determined precisely, but could be as high as $10^6$ cm$^{-3}$ based on collisional excitation requirements for some molecular transitions observed (Sakai *et al.* 2009). The molecular cloud L1495B has been studied by Cordiner et al. (2013), who derive a volume density of $1.1 \times 10^4$ cm$^{-3}$ and estimate a kinetic temperature of 10 K. The column density of $H_2$ in L1495B is $1.2 \times 10^{22}$ cm$^{-2}$ (Myers *et al.* 1983). In the core L1521F, the density of particles is $1.1 \times 10^6$ cm$^{-3}$, the rotational temperature of $N_2H^+$ of 4.8 K is consistent with a gas kinetic temperature of 10 K, and $N(H_2)$ is $1.35 \times 10^{23}$ cm$^{-2}$ (Crapsi *et al.* 2005). The last targeted source is the clump 1a of the Serpens South complex, which has been studied by Friesen et al. (2013). These authors detect abundant cyanopolyynes in some clumps and derive a kinetic temperature of 11 K through the observation of $NH_3$ lines, suggesting a volume density of $\sim 10^4$ cm$^{-3}$. The column density of $H_2$ is $2 \times 10^{22}$ cm$^{-2}$.

Propynal was detected in the seven targeted sources through the $9_{0,9}$-$8_{0,8}$ rotational transition lying at 83.8 GHz. The detection of this line in B1-b was previously presented by Cernicharo et al. (2012). An additional transition of propynal, $5_{1,5}$-$4_{0,4}$ at 107.5 GHz, was also detected in TMC-1 (see Table 2). This line was not detected in B1-b, probably due to an insufficient sensitivity, while in the rest of sources this frequency was not covered by the observations. We detected cyclopropenone in four of the seven observed sources, in B1-b, L483, and Lupus-1A through the $7_{1,6}$-$6_{1,5}$ ortho transition at 103.1 GHz. In TMC-1 this transition was not clearly detected due to an insufficient sensitivity, although another rotational transition belonging also to the ortho species of cyclopropenone, the $7_{1,7}$-$6_{1,6}$ lying at 92.5 GHz, could be tentatively detected (see Figure 1 and Table 2). The third isomer, propadienone, was not detected in any of the seven targeted sources. As already noted, the bent structure of propadienone leads to a complex and dense rotational spectrum which does not favor its detection. The most stringent upper limits to the column density of propadienone are set by the non detection of the $10_{0,10}$ (0⁻)-$9_{0,9}$ (0⁻) transition of the ortho species, which lies at 86.4 GHz and has an upper level energy of 22.8 K with respect to the ortho ground state. Transition frequencies and quantum numbers of the observed lines are listed in Table 2.

Table 2. Observed line parameters.

| Source | Molecule | Transition | Frequency (MHz) | $E_{up}$ (K) | $A_{ul}$ (s$^{-1}$) | HPBW (″) | $B_{eff}/F_{eff}$ | $V_{LSR}$ (km s$^{-1}$) | $\Delta v$ (km s$^{-1}$) | $\int T_A^* dv$ (K km s$^{-1}$) |
|---|---|---|---|---|---|---|---|---|---|---|
| TMC-1 | HCCCHO | $9_{0,9} - 8_{0,8}$ | 83775.832 | 20.1 | $1.80 \times 10^{-5}$ | 29.0 | 0.86 | +5.72(5) | 0.39(13) | 0.006(2) |
|  | HCCCHO | $5_{1,5} - 4_{0,4}$ | 107556.726 | 9.6 | $8.58 \times 10^{-6}$ | 22.6 | 0.83 | +5.86(4) | 0.48(8) | 0.007(1) |
|  | c-C$_3$H$_2$O | $7_{1,7} - 6_{1,6}$ | 92517.434 | 17.2 | $8.11 \times 10^{-5}$ | 26.3 | 0.85 | +5.65(6) | 0.43(13) | 0.004(1) |
| B1-b | HCCCHO | $9_{0,9} - 8_{0,8}$ | 83775.832 | 20.1 | $1.80 \times 10^{-5}$ | 29.0 | 0.86 | +6.43(6) | 0.60(12) | 0.012(2) |
|  | c-C$_3$H$_2$O | $7_{1,6} - 6_{1,5}$ | 103069.924 | 19.2 | $1.12 \times 10^{-4}$ | 23.6 | 0.84 | +6.49(4) | 0.44(9) | 0.006(1) |
| L483 | HCCCHO | $9_{0,9} - 8_{0,8}$ | 83775.832 | 20.1 | $1.80 \times 10^{-5}$ | 29.0 | 0.86 | +5.24(4) | 0.34(12) | 0.009(2) |
|  | c-C$_3$H$_2$O | $7_{1,6} - 6_{1,5}$ | 103069.924 | 19.2 | $1.12 \times 10^{-4}$ | 23.6 | 0.84 | +5.27(3) | 0.26(6) | 0.008(2) |
| Lupus-1A | HCCCHO | $9_{0,9} - 8_{0,8}$ | 83775.832 | 20.1 | $1.80 \times 10^{-5}$ | 29.0 | 0.86 | +5.03(4) | 0.38(10) | 0.007(1) |
|  | c-C$_3$H$_2$O | $7_{1,6} - 6_{1,5}$ | 103069.924 | 19.2 | $1.12 \times 10^{-4}$ | 23.6 | 0.84 | +4.98(4) | 0.34(10) | 0.005(1) |
| L1495B | HCCCHO | $9_{0,9} - 8_{0,8}$ | 83775.832 | 20.1 | $1.80 \times 10^{-5}$ | 29.0 | 0.86 | +7.60(4) | 0.25(5) | 0.005(1) |
| L1521F | HCCCHO | $9_{0,9} - 8_{0,8}$ | 83775.832 | 20.1 | $1.80 \times 10^{-5}$ | 29.0 | 0.86 | +6.38(6) | 0.46(11) | 0.005(1) |
| Serpens South 1a | HCCCHO | $9_{0,9} - 8_{0,8}$ | 83775.832 | 20.1 | $1.80 \times 10^{-5}$ | 29.0 | 0.86 | +7.51(5) | 0.74(13) | 0.020(3) |

Numbers in parentheses are 1$\sigma$ uncertainties in units of the last digits. Antenna temperature ($T_A^*$) can be converted to main beam brightness temperature ($T_{MB}$) by dividing by ($B_{eff}/F_{eff}$).

Column density values or upper limits for the three isomers of H$_2$C$_3$O were derived in the seven observed sources assuming local thermodynamic equilibrium and a rotational temperature of 10 K. These values are given in Table 3 as well as the physical conditions corresponding to the various sources.

Table 3. Column densities and fractional abundances relative to H$_2$.

| Source | $n$(H$_2$) (cm$^{-3}$) | $N$(H$_2$) (cm$^{-2}$) | $N$(HCCCHO) (cm$^{-2}$) | $N$(c-C$_3$H$_2$O) (cm$^{-2}$) | $N$(H$_2$C$_3$O) (cm$^{-2}$) | $f$(HCCCHO) | $f$(c-C$_3$H$_2$O) | $f$(H$_2$C$_3$O) |
|---|---|---|---|---|---|---|---|---|
| TMC-1 | $(2-8) \times 10^4$ | $1.0 \times 10^{22}$ | $8.0 \times 10^{11}$ | $5.4 \times 10^{10}$ | $< 2.1 \times 10^{11}$ | $8.0 \times 10^{-11}$ | $5.4 \times 10^{-12}$ | $< 2.1 \times 10^{-11}$ |
| B1-b | $1 \times 10^5$ | $1.5 \times 10^{23}$ | $7.9 \times 10^{11}$ | $8.9 \times 10^{10}$ | $< 2.9 \times 10^{11}$ | $5.3 \times 10^{-12}$ | $5.9 \times 10^{-13}$ | $< 1.9 \times 10^{-12}$ |
| L483 | $3.4 \times 10^4$ | $3.0 \times 10^{22}$ | $6.0 \times 10^{11}$ | $1.2 \times 10^{11}$ | $< 2.3 \times 10^{11}$ | $2.0 \times 10^{-11}$ | $4.0 \times 10^{-12}$ | $< 7.7 \times 10^{-12}$ |
| Lupus-1A | $(1-10) \times 10^5$ | $1.5 \times 10^{22}$ | $4.6 \times 10^{11}$ | $7.4 \times 10^{10}$ | $< 2.1 \times 10^{11}$ | $3.1 \times 10^{-11}$ | $4.9 \times 10^{-12}$ | $< 1.4 \times 10^{-11}$ |
| L1495B | $1.1 \times 10^4$ | $1.2 \times 10^{22}$ | $3.3 \times 10^{11}$ | $< 2.9 \times 10^{10}$ | $< 1.4 \times 10^{11}$ | $2.8 \times 10^{-11}$ | $< 2.4 \times 10^{-12}$ | $< 1.2 \times 10^{-11}$ |
| L1521F | $1.1 \times 10^6$ | $1.35 \times 10^{23}$ | $3.3 \times 10^{11}$ | $< 4.2 \times 10^{10}$ | $< 1.8 \times 10^{11}$ | $2.4 \times 10^{-12}$ | $< 3.1 \times 10^{-13}$ | $< 1.3 \times 10^{-12}$ |
| Serpens South 1a | $1.0 \times 10^4$ | $2.0 \times 10^{22}$ | $1.3 \times 10^{12}$ | $< 8.2 \times 10^{10}$ | $< 2.9 \times 10^{11}$ | $6.5 \times 10^{-11}$ | $< 4.1 \times 10^{-12}$ | $< 1.5 \times 10^{-11}$ |

Column densities are derived adopting a rotational temperature of 10 K. Upper limits to the column densities are given at a 3$\sigma$ confidence level.

In all the sources the most abundant isomer is propynal, while cyclopropenone is 5 to >16 times less abundant and propadienone is also less abundant than propynal by a factor of at least 2-4. It should be noted that the upper limits placed on the column density of propadienone are not very stringent and therefore do not allow us to conclude which isomer, cyclopropenone or propadienone, is more abundant in cold dark clouds. It is interesting to note that in Sagittarius B2, Loomis et al. (2015) also found that propynal is the most abundant isomer, while cyclopropenone is 10 times less abundant and propadienone, which was not detected, is more than 100 times less abundant.

3 Discussion

3.1. The chemical model

The chemical reactions for the various $H_2C_3O$ isomers formation and destruction listed in Table 4 have been implemented in the chemical model Nautilus (Hersant *et al.* 2009, Semenov *et al.* 2010). The Nautilus code computes the gas-phase and grain surface icy composition as a function of time taking into account reactions in the gas-phase, physisorption onto and desorption from grain surfaces and reactions at the surface of the grains. The surface reactions are based on the Langmuir-Hinshelwood mechanism with the formalism of Hasegawa *et al.* (1992). For desorption, we consider thermal desorption as well as desorption induced by cosmic-rays (Hasegawa & Herbst 1993) and exothermic chemical reactions (Garrod et al. 2007). The gas-phase network is based on kida.uva.2014[2] (Wakelam et al. 2015), with the modifications described in section 3.2. The surface network and parameters are the same as in Ruaud et al (2015), i.e. it includes low temperature Eley-Rideal and complex induced reaction mechanisms.

The chemical composition of the gas-phase and grain surfaces is computed as a function of time. The gas and dust temperatures are set equal to 10 K, the total H density is varied between $2\times10^4$ cm$^{-3}$ and $2\times10^6$ cm$^{-3}$ for molecular clouds, to study the of the density. The cosmic-ray ionization rate is equal to $1.3\times10^{-17}$ s$^{-1}$ and the visual extinction is equal to 30. All elements are assumed to be initially in atomic form, except for hydrogen, which is entirely molecular. The initial abundances are similar to those of Table 1 of Hincelin *et al.* (2011), the C/O elemental ratio being equal to 0.7 in this study (the C/O ratio has been varied over the range 0.7 – 1.0, inducing only small modifications in the abundances of the $H_2C_3O$ isomers).

3.2. $H_2C_3O$ chemistry

To describe $H_2C_3O$ chemistry we need to introduce the three most stable isomers (propadienone ($H_2C=C=C=O$), cyclopropenone (c-$C_3H_2O$) and propynal (HCCCHO) (all other isomers being unstable or much less stable)), the $HC_3O$ (HCCCO) species, and the three isomers of $H_3C_3O^+$ ($HCCCHOH^+$, $c$-$C_3H_2OH^+$ and $C_2H_3CO^+$). The structures and the energies of the introduced species have been calculated at the DFT level. To construct the chemical scheme to describe $H_2C_3O$ isomers we performed a bibliographic review of previous work and we also considered other potentially important formation and loss processes. The presence of any potential barrier for the most important reactions has also been studied at the DFT level. The reactions used to describe the chemistry of $H_2C_3O$ isomers are listed in Table 4, most of them being new reactions in the network.

---

[2] http://kida.obs.u-bordeaux1.fr/models

Some formation routes for the $H_2C_3O$ isomers were proposed by Petrie (1995). Petrie noted, following Scott et al (1995) and Mclagan et al (1995), that electronic dissociative recombination of the various $C_2H_3CO^+$ ions, formed through the $C_2H_3^+$ + CO reaction (Herbst *et al.* 1984), was unlikely to be an efficient way to form $H_2C_3O$ isomers because, if electronic dissociative recombination of the various $H_3C_3O^+$ isomers did preserve the carbon skeleton, the most stable $H_2C_3O$ isomer, propadienone ($H_2C=C=C=O$), should be abundant in molecular clouds. Instead, he proposed that propynal could be formed through the $C_2H$ + $H_2CO$ reaction, cyclopropenone through the OH + $c$-$C_3H_2$ reaction, and propadienone through the OH + $l$-$C_3H_2$ reaction. However he did not characterize the entrance valley energetics for these reactions. More recent calculations (Dong *et al.* 2005) have shown that the $C_2H$ + $H_2CO$ → $C_2H_2$ + HCO reaction does not have any barrier but that the C addition channel, $C_2H$ + $H_2CO$ → HCCCHO + H, presents a small barrier calculated between 9 and 15 kJ/mol at a relatively high level of calculation (CCSD(T)/6-311+G). Consequently, the $C_2H$ + $H_2CO$ reaction should be included in the network but it is not a formation route for propynal in the cold conditions of interstellar clouds. Concerning the OH + $c$-$C_3H_2$ reaction suggested by Petrie (1995), we performed DFT calculations showing that this reaction proceeds without a barrier, leading to cyclopropenone, which is the most kinetically favored spin allowed channel. Moreover, the OH + $l$-$C_3H_2$ reaction is very likely to form propadienone without a barrier and the OH + $t$-$C_3H_2$ reaction produces propynal ($t$-$C_3H_2$ is the third $C_3H_2$ isomer, with a stability in between the cyclic and the linear ones (Aguilera-Iparraguirre *et al.* 2008, Vazquez *et al.* 2009)). $l$-$C_3H_2$ is about 10 times less abundant than $c$-$C_3H_2$ in dense molecular clouds such as TMC-1 (Fossé *et al.* 2001), in good agreement with our model (see Sect. 3.3). As a result, the OH + $l$-$C_3H_2$ reaction involves much smaller fluxes in dark clouds than the OH + $c$-$C_3H_2$ reaction. There are no observations of $t$-$C_3H_2$, which has no, or very low, permanent dipole moment, but our model predicts similar amounts of $t$-$C_3H_2$ and $l$-$C_3H_2$. Consequently, the OH + $t$-$C_3H_2$ reaction should also involve small fluxes. The efficiency of OH + $C_3H_2$ reactions to produce $H_2C_3O$ isomers is linked to the fact that this channel is a direct OH addition followed by H elimination associated to the fact that CO production in its ground state is spin forbidden and also to the fact that HOC radical production, which involve also few steps, has a very low stability and then the $C_2H_2$ + HOC channel is kinetically unfavored versus $H_2C_3O$ + H production. Ahmadvand *et al.* (2014) proposed three reactions to produce cyclopropenone, O + c-$C_3H_2$, $O_2$ + c-$C_3H_2$ and CO + $C_2H_2$ reactions. Among them, the O + c-$C_3H_2$ reaction has an open bimolecular exit channel so that the association reaction will be negligible and the CO + $C_2H_2$ reaction shows a very high barrier so it will also be negligible. The third one, the $O_2$ + c-$C_3H_2$ reaction is found to be without a barrier using DFT (B3LYP) and Completely Renormalized Coupled Cluster (CR-CCL) methods. This reaction may be an efficient way to produce cyclopropenone. However we have performed additional DFT calculations and have found a transition state for this reaction. To describe more accurately the chemistry of $H_2C_3O$ isomers we searched for additional reactions, which might potentially be involved. We find that the most interesting one is the O + $C_3H_3$ reaction. The rate constant for the O + $C_3H_3$ reaction has been determined experimentally between 295 and 750 K to be equal to k = $2.3\times10^{-10}$ $cm^3$ $molecule^{-1}$ $s^{-1}$ (Slagle *et al.* 1991), characteristic of a barrierless reaction. Consequently, the rate constant at low temperature might be similar to those of the O + alkene reaction rate constants studied between 25 and 500 K (Sabbah *et al.* 2007). Slagle et al. (1991) detected $H_2C_3O$ as the main product for the O + $C_3H_3$ reaction, in good agreement with combined crossed-beam and *ab initio* investigations showing that the main exit channel is H + $H_2C_3O$ with minor channels leading to $C_3H_2$ + OH, $C_2H_2$ + HCO and $C_2H_3$ + CO (Kwon *et al.* 2006, Lee *et al.* 2003, Lee *et al.* 2006). RRKM calculations show that among the various $H_2C_3O$ isomers, propynal is favored (Kwon et al. 2006, Lee et al. 2006) with very little, if any, propadienone production. None of these papers specifically studied cyclopropenone production but considering the ab-initio potential-energy surface, it is clearly an unfavorable exit channel. The O + $C_3H_3$ reaction is then unlikely to represent a way to produce cyclopropenone. Moreover, the absence of detection for propadienone in the ISM is in good agreement with theoretical and experimental results showing that the O + $C_3H_3$ reaction produces mainly propynal.

To summarize, our model suggests that propynal is produced through the O + $C_3H_3$ reaction, cyclopropenone is produced through the OH + c-$C_3H_2$ reaction, and propadienone is produced through the OH + l-$C_3H_2$ reaction, although this last reaction involves a small flux. In our present model, no surface reactions directly produce $H_2C_3O$, whereas c,l-$C_3H_2$ and $C_3H_3$ are mainly produced through gas-phase reactions (c,l-$C_3H_3^+$ dissociative electronic recombination for c,l,t-$C_3H_2$ production, with minor contributions from the CH + $C_2H_2$, and C + $C_2H_4$ reactions for $C_3H_3$ production (Haider & Husain 1993, Chastaing *et al.* 1999, Chastaing *et al.* 2001, Bergeat & Loison 2001)).

The three $H_2C_3O$ isomers are all lost mainly through their reactions with carbon atoms (even if there are no experimental or theoretical data on the C + $H_2C_3O$ reactions, we can deduce the absence of barrier from similar systems such as the C + $H_2CO$ (Husain & Ioannou 1999), C + $CH_3CHO$ (Husain & Ioannou 1999) and C + $CH_3OH$ (Shannon *et al.* 2014) reactions). The three $H_2C_3O$ isomers will also be lost through protonation (by their reactions with $H_3^+$ and $HCO^+$) followed by electronic dissociative recombination which does not give back $H_2C_3O$ isomers (or with a very low branching ratio) (Petrie 1995, Scott et al. 1995, Maclagan et al. 1995). Reactions of $H_2C_3O$ isomers with OH radicals are minor loss processes. In this study we do not consider the reactions of $H_2C_3O$ isomers with CH, $C_2$, $C_2H$ and CN radicals, which should proceed without barriers, but involving much lower fluxes.

It should be noted that the chemistry of $H_2C_3O$ isomers involves relatively few reactions most of them being gas-phase reactions only.

Table 4: Summary of reaction review.
$H_2CCCO$ = propadienone, HCCCHO = propynal , $c-C_3H_2O$ = cyclopropenone
$k = \alpha \times (T/300)^\beta \times \exp(-\gamma/T)$ cm$^3$ molecule$^{-1}$ s$^{-1}$ , T range is 10-300K
Ionpol1: $k = \alpha\beta(0.62+0.4767\gamma (300/T)^{0.5})$ cm$^3$ molecule$^{-1}$ s$^{-1}$ , (Wakelam *et al.* 2012, Wakelam *et al.* 2010)
$F_0 = \exp(\Delta k/k_0)$ and $F(T)=F_0\times\exp(g\times|1/T-1/T_0|)$

| | Reaction | | α | β | γ | $F_0$ | g | ref |
|---|---|---|---|---|---|---|---|---|
| 1. | $C^+ + H_2CCCO$ | → $c-C_3H_2^+$ + CO | 2.0e-9 | 0 | 0 | 2 | 0 | / $C^+ + H_2CO$ (Anicich 2003) |
| | | → $l-C_3H_2^+$ + CO | 1.0e-9 | 0 | 0 | 2 | 0 | |
| | | → $H_2C_3O^+$ + C | 1.0e-9 | 0 | 0 | 2 | 0 | |
| 2. | $C^+ + c-C_3H_2O$ | → $c-C_3H_2^+$ + CO | 2.0e-9 | 0 | 0 | 2 | 0 | / $C^+ + H_2CO$ (Anicich 2003) |
| | | → $l-C_3H_2^+$ + CO | 1.0e-9 | 0 | 0 | 2 | 0 | |
| | | → $H_2C_3O^+$ + C | 1.0e-9 | 0 | 0 | 2 | 0 | |
| 3. | $C^+$ + HCCCHO | → $c-C_3H_2^+$ + CO | 2.0e-9 | 0 | 0 | 2 | 0 | / $C^+ + H_2CO$ (Anicich 2003) |
| | | → $l-C_3H_2^+$ + CO | 1.0e-9 | 0 | 0 | 2 | 0 | |
| | | → $H_2C_3O^+$ + C | 1.0e-9 | 0 | 0 | 2 | 0 | |
| 4. | C + HCCCHO | → $t-C_3H_2$ + CO | 2.0e-10 | 0 | 0 | 2 | 0 | Rate constant / C + $H_2CO$, C + $CH_3CHO$ (Husain & Ioannou 1999) and C + $C_2H_4$ (Chastaing *et al.* 1999, Chastaing *et al.* 2001, Bergeat & Loison 2001, Haider & Husain 1993a, Haider & Husain 1993b). Simplified products $t-C_3H_2$ is the third isomer of $C_3H_2$ with a stability between $c-C_3H_2$ and $l-C_3H_2$ (Vazquez *et al.* 2009, Aguilera-Iparraguirre *et al.* 2008) |
| 5. | C + $c-C_3H_2O$ | → $c-C_3H_2$ + CO | 2.0e-10 | 0 | 0 | 2 | 0 | Rate constant / C + $H_2CO$, C + $CH_3CHO$ (Husain & Ioannou 1999) and C + $C_2H_4$ (Chastaing *et al.* 1999, Chastaing *et al.* 2001, Bergeat & Loison 2001, Haider & Husain 1993a, Haider & Husain 1993b). Simplified products |
| 6. | C + $H_2CCCO$ | → $l-C_3H_2$ + CO | 2.0e-10 | 0 | 0 | 2 | 0 | Rate constant / C + $H_2CO$, C + $CH_3CHO$ (Husain & Ioannou 1999) and C + $C_2H_4$ (Chastaing *et al.* 1999, Chastaing *et al.* 2001, Bergeat & Loison 2001, Haider & Husain 1993a, Haider & Husain 1993b). Simplified products |
| 7. | O + $c-C_3H_2$ | → HCCCO + H | 4.0e-11 | 0 | 0 | 2 | 0 | No barrier at M06, MP2 and CCSD level. $C_2H$ + CO branching ratio from (Boullart *et al.* 1996) |
| | | → HCO + $C_2H$ | 1.0e-11 | 0 | 0 | 2 | 0 | |
| | | → H + CO + $C_2H$ | 1.0e-11 | 0 | 0 | 2 | 0 | |
| | | → CO + $C_2H_2$ | 3.0e-11 | 0 | 0 | 2 | 0 | |
| 8. | O + $l-C_3H_2$ | → HCCCO + H | 2.0e-11 | 0 | 0 | 2 | 0 | No barrier at M06, MP2 and CCSD level. $C_2H$ + CO branching ratio from (Boullart et al. 1996) |
| | | → H + CO + $C_2H$ | 4.0e-11 | 0 | 0 | 2 | 0 | |
| | | → CO + $C_2H$ | 4.0e-11 | 0 | 0 | 2 | 0 | |
| 9. | O + $t-C_3H_2$ | → HCCCO + H | 3.0e-11 | 0 | 0 | 2 | 0 | / O + $^3CH_2$ (Bohland *et al.* 1984, Vinckier & Debruyn 1979) |
| | | → HCO + $C_2H$ | 1.0e-11 | 0 | 0 | 2 | 0 | |
| | | → H + CO + $C_2H$ | 1.0e-11 | 0 | 0 | 2 | 0 | |

| # | Reaction | | Rate | α | β | γ | F | g | Comments |
|---|---|---|---|---|---|---|---|---|---|
| 10. | O + C$_3$H$_3$ | → HCCCHO + H | 1.6e-10 | 0 | 0 | 2 | 0 | | Global rate constant from (Slagle *et al.* 1991), branching ratio deduced from (Lee *et al.* 2006). |
| | | → C$_2$H$_2$ + HCO | 2.0e-11 | 0 | 0 | 2 | 0 | | |
| | | → C$_2$H$_3$ + CO | 5.0e-11 | 0 | 0 | 2 | 0 | | |
| | | → c-C$_3$H$_2$ + OH | 1.0e-11 | 0 | 0 | 2 | 0 | | |
| 11. | OH + c-C$_3$H$_2$ | → c-C$_3$H$_2$O + H | 2.0e-10 | 0 | 0 | 3 | 0 | | No barrier for $^2$c-C$_3$H$_2$OH formation, high exit barrier for $^2$H + $^1$CO + $^1$C$_2$H$_2$. |
| | | → H + CO + C$_2$H$_2$ | 0 | 0 | 0 | 0 | 0 | | |
| 12. | OH + l-C$_3$H$_2$ | → H$_2$CCCO + H | 2.0e-10 | 0 | 0 | 3 | 0 | | / OH + c-C$_3$H$_2$ |
| | | → H + CO + C$_2$H$_2$ | 0 | 0 | 0 | 0 | 0 | | |
| 13. | OH + t-C$_3$H$_2$ | → HCCCHO + H | 1.0e-10 | 0 | 0 | 3 | 0 | | / OH + c-C$_3$H$_2$ |
| | | → H + CO + C$_2$H$_2$ | 0 | 0 | 0 | 0 | 0 | | |
| 14. | OH + HCCCHO | → HCCCO + H$_2$O | 1.0e-11 | -0.6 | 0 | 2 | 0 | | Simplified, but realistic, products |
| | | → H$_2$CCO + HCO | 1.0e-11 | -0.6 | 0 | 2 | 0 | | |
| 15. | OH + c-C$_3$H$_2$O | → H$_2$CCO + HCO | 2.0e-11 | -0.6 | 0 | 5 | 0 | | Oversimplified products. Rate constant may be much lower. |
| 16. | OH + H$_2$CCCO | → CH$_3$CO + CO | 2.0e-11 | -0.6 | 0 | 2 | 0 | | Oversimplified products |
| 17. | C$_2$H$_3^+$ + CO | → C$_2$H$_3$CO$^+$ | 2.0e-15 | -2.5 | 0 | 10 | 0 | | (Herbst *et al.* 1984) deduced the radiative association rate constant from (unpublished) termolecular rate. (Scott *et al.* 1995, Maclagan *et al.* 1995) identified the product of the reaction as H$_2$C=CH-CO$^+$, the most stable C$_3$H$_3$O$^+$ isomer. |
| 18. | HCCCHO + H$_3^+$ | → HCCCHOH$^+$ + H$_2$ | 1.0 | 3.2e-9 | 4.1 | 3 | 0 | | Capture rate, Ionpol1 |
| 19. | HCCCHO + HCO$^+$ | → HCCCHOH$^+$ + CO | 1.0 | 1.2e-9 | 4.1 | 3 | 0 | | Capture rate, Ionpol1 |
| 20. | HCCCHO + H$_3$O$^+$ | → HCCCHOH$^+$ + H$_2$O | 1.0 | 1.4e-9 | 4.1 | 3 | 0 | | Capture rate, Ionpol1 |
| 21. | c-C$_3$H$_2$O + H$_3^+$ | → c-C$_3$H$_2$OH$^+$ + H$_2$ | 0.5 | 3.2e-9 | 6.3 | 3 | 0 | | Capture rate, Ionpol1 |
| | | → C$_2$H$_3$CO$^+$ + H$_2$ | 0.5 | 3.2e-9 | 6.3 | 3 | 0 | | |
| 22. | c-C$_3$H$_2$O + HCO$^+$ | → c-C$_3$H$_2$OH$^+$ + CO | 0.5 | 1.2e-9 | 6.3 | 3 | 0 | | Capture rate, Ionpol1 |
| | | → C$_2$H$_3$CO$^+$ + CO | 0.5 | 1.2e-9 | 6.3 | 3 | 0 | | |
| 23. | c-C$_3$H$_2$O + H$_3$O$^+$ | → c-C$_3$H$_2$OH$^+$ + H$_2$O | 0.5 | 1.4e-9 | 6.3 | 3 | 0 | | Capture rate, Ionpol1 |
| | | → C$_2$H$_3$CO$^+$ + H$_2$O | 0.5 | 1.4e-9 | 6.3 | 3 | 0 | | |
| 24. | H$_2$CCCO + H$_3^+$ | → C$_2$H$_3$CO$^+$ + H$_2$ | 1.0 | 3.2e-9 | 3.8 | 3 | 0 | | Capture rate, Ionpol1 |
| 25. | H$_2$CCCO + HCO$^+$ | → C$_2$H$_3$CO$^+$ + CO | 1.0 | 1.2e-9 | 3.8 | 3 | 0 | | Capture rate, Ionpol1 |
| 26. | H$_2$CCCO + H$_3$O$^+$ | → C$_2$H$_3$CO$^+$ + H$_2$O | 1.0 | 1.4e-9 | 3.8 | 3 | 0 | | Capture rate, Ionpol1 |
| 27. | H$_2$C$_3$O$^+$ + e$^-$ | → H + HC$_3$O | 1.0e-7 | -0.7 | 0 | 3 | 0 | | By comparison with similar reactions (Florescu-Mitchell & Mitchell 2006, Fournier *et al.* 2013), branching ratio roughly guessed |
| | | → C$_2$H$_2$ + CO | 2.0e-7 | -0.7 | 0 | 3 | 0 | | |
| | | → C$_2$H + H + CO | 4.0e-7 | -0.7 | 0 | 3 | 0 | | |
| | | → H + H + C$_3$O | 1.0e-7 | -0.7 | 0 | 3 | 0 | | |

| # | Reaction | | Rate | β | γ | F | g | Reference/Comment |
|---|---|---|---|---|---|---|---|---|
| | | → $H_2 + C_3O$ | | | | | | |
| 28. | $HCCCHOH^+ + e^-$ | → $H + HCCCHO$ | 4.0e-8 | -0.7 | 0 | 3 | 0 | By comparison with similar reactions (Florescu-Mitchell & Mitchell 2006, Fournier et al. 2013), branching ratio roughly guessed |
| | | → $H + H + HC_3O$ | 2.0e-7 | -0.7 | 0 | 10 | 0 | |
| | | → $H + C_2H_2 + CO$ | 6.0e-7 | -0.7 | 0 | 3 | 0 | |
| 29. | $c-C_3H_2OH^+ + e^-$ | → $H + c-C_3H_2O$ | 4.0e-8 | -0.7 | 0 | 3 | 0 | By comparison with similar reactions (Florescu-Mitchell & Mitchell 2006, Fournier et al. 2013), branching ratio roughly guessed |
| | | → $H + H + HC_3O$ | 1.0e-7 | -0.7 | 0 | 3 | 0 | |
| | | → $H + C_2H_2 + CO$ | 7.0e-7 | -0.7 | 0 | 3 | 0 | |
| 30. | $C_2H_3CO^+ + e^-$ | → $H + H_2C_3O$ | 4.0e-8 | -0.7 | 0 | 3 | 0 | By comparison with similar reactions (Florescu-Mitchell & Mitchell 2006, Fournier et al. 2013). Branching ratio deduced from the fact that $H_2C_3O$ is not detected. |
| | | → $H + H + HC_3O$ | 4.0e-8 | -0.7 | 0 | 3 | 0 | |
| | | → $H + C_2H_2 + CO$ | 8.0e-7 | -0.7 | 0 | 3 | 0 | |

3.3. Comparison with observations

Figure 2 shows the results of the simulation for the various $H_2C_3O$ isomers for typical molecular cloud conditions ($n(H_2) = 2\times10^4$ cm$^{-3}$). Figure 3 shows the results of the simulation for typical starless cores with higher densities ($n(H_2) = 2\times10^5$ cm$^{-3}$).

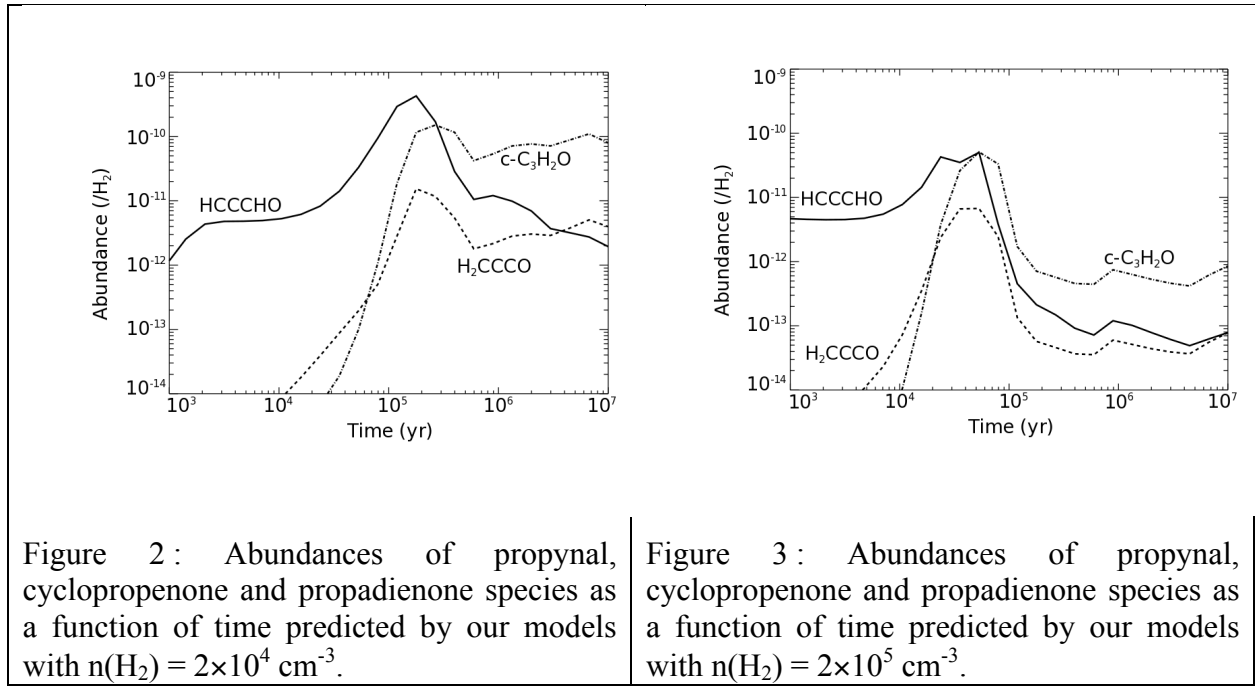

Figure 2 : Abundances of propynal, cyclopropenone and propadienone species as a function of time predicted by our models with $n(H_2) = 2\times10^4$ cm$^{-3}$.

Figure 3 : Abundances of propynal, cyclopropenone and propadienone species as a function of time predicted by our models with $n(H_2) = 2\times10^5$ cm$^{-3}$.

The comparisons with the observations are delicate due to the large variations and uncertainties in the densities of the interstellar objects observed, in the range $10^4$-$10^6$ cm$^{-3}$. We consider two different types of clouds among the objects studied here. Firstly, we simulate a typical dense cloud with a value for $n(H_2)$ of a few $10^4$ cm$^{-3}$, which corresponds to objects such as TMC-1, L483, L1495B and Serpens South 1a. Secondly, we simulate an evolved cloud with a higher density with a value for $n(H_2)$ of a few $10^5$ cm$^{-3}$, corresponding to the objects L1521F and B1 clouds. For the starless core Lupus-1A, the volume density has not been determined precisely. The density value of $10^6$ cm$^{-3}$ based on collisional excitation requirements for some observed molecular transitions (Sakai *et al.* 2009) is not coherent with our calculated abundances for the species studied here (see below). All the sources have a similar kinetic temperature around 10 K and a high visual extinction above 10 mag. The most abundant isomer observed is propynal. In molecular clouds having a density of a few $10^4$ cm$^{-3}$, TMC-1, L483, L1495B and Serpens South 1a, the best agreement between calculations and observations for propynal is for a cloud age around $10^5$ years. Increasing the simulated total density leads to an acceleration of the chemistry (the maximum of the HCCCHO abundance is moved to shorter time for example) associated to a slight decrease of the abundance, in relatively good agreement with observations for L1521F and B1. These effects are amplified when we increase the simulated total density up to $10^6$ cm$^{-3}$. The agreement is less good for Lupus 1A considering high tiotal density. However the density of this object is highly uncertain and the agreement is better for Lupus 1A if its density is considered to be closer to the lower estimate (n(H2) around 1e5 cm-3) or even smaller. Considering the uncertainties of the observations, particularly with respect to the density determination, the agreement is good between propynal observations and models for a relatively early age of various clouds. Cyclopropenone is detected in some, but not all, clouds, the upper limit in some cases being quite stringent. The observations in TMC-1 and L483 are, as for propynal, in good agreement for an early age around $10^5$ years. In that case the calculated propynal/cyclopropenone ratio, equal to 15 and 5 respectively, are also in good agreement with observations. As cyclopropenone is mainly produced through the c-$C_3H_2$ + OH reaction in our model, it is interesting to compare the simulated and observed abundances for this species. In a typical dense molecular cloud such as TMC- 1, c-$C_3H_2$ is measured to be equal to $5.8 \times 10^{-9} \times [H_2]$ (Fossé et al. 2001), in very good agreement with our calculations for a cloud age around $10^5$ years and a density of $n(H_2) = 2 \times 10^4$ cm$^{-3}$ as shown in Figure 4.

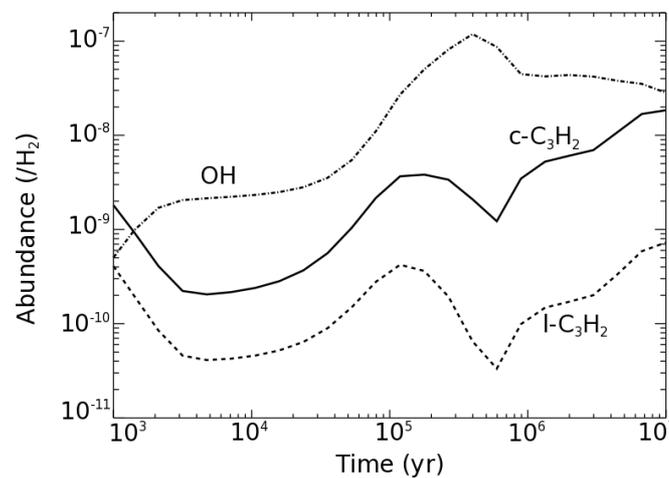

Figure 4 : Abundances of c-$C_3H_2$ (continuous lines), l-$C_3H_2$ species (dashed lines) and OH (dash-dot line)as a function of time predicted by our models with $n(H_2) = 2 \times 10^4$ cm$^{-3}$.

The combined good agreement for TMC-1 between observations and calculations for c-$C_3H_2O$ (this work), c-$C_3H_2$ (Fossé et al. 2001) and OH ($1.0 \times 10^{-7} \times [H_2]$) (Suutarinen et al. 2011) is an indicator of the reliability of our chemical model. It is worth noting that observations of l-$C_3H_2$ in TMC-1 leading to an abundance of $2.1 \times 10^{-10} \times [H_2]$ (Fossé et al. 2001), is also in good agreement with our calculations for a cloud age around $10^5$ years and a density of $n(H_2) = 2 \times 10^4$ cm$^{-3}$ (see Figure 4). Then the calculated propadienone (produced mainly through l-$C_3H_2$ + OH reaction) abundance is likely to be reliable and propadienone may be detectable in TMC-1 or in similar clouds.

At large evolution times, the simulated propynal abundance shows a notable decay (the $C_3H_3$ abundance falls, so that the O + $C_3H_3$ → propynal + H reaction is much less efficient). This is not the case for cyclopropenone as in our model, c-$C_3H_2$ is still abundant at large evolution times). These two effects are uncorrelated, resulting from separate processes in the network.

It is worth noting that the most stable isomer, propadienone, is not the most abundant isomer, its upper limit being notably lower than the detected abundance of propynal. This result shows that the « minimum energy principle », which states that the most energetically stable isomer should be the most abundant (Lattelais et al. 2009), is questionable in interstellar conditions. This is due to the fact that species are controlled by their kinetics of formation and not through thermodynamic principles, the reaction(s) producing propadienone being less efficient that those producing propynal and cyclopropenone. Despite the unfavorable production of propadienone, the calculated relative abundance (compared to $H_2$) for a typical cloud age (between $10^5$ and $10^6$ years) is between $10^{-11}$ and $10^{-12}$ and considering its electric dipole moment (2.5 Debye at DFT level), propadienone may still be detectable in some regions despite its densely populated rotational spectrum.

4. Conclusions

In this study, we report the detection of two $H_2C_3O$ isomers, propynal (HC≡C-CH=O) toward seven cold dark clouds and cyclopropenone (c-$C_3H_2O$) toward four cold dark clouds using the IRAM 30m telescope, as well as the non-detection of a third isomer, propadienone ($H_2C=C=C=O$). It is interesting to note that the most abundant isomer is not the most stable one, propadienone, but propynal.

To understand the chemistry of these compounds we have reviewed the formation and loss of the three $H_2C_3O$ isomers. The chemical network to describe $H_2C_3O$ isomers has been reviewed in detail and preliminary DFT calculations have been performed on key reactions. Cyclopropenone and propadienone were not present in the kida.uva.2014 network. We have shown that the chemistry of these species involves mainly gas-phase reactions and that the abundances of $H_2C_3O$ isomers are controlled by the kinetics and not by the thermodynamics, as already emphasized by Loomis et al. 2015. A key point of the model is, following Petrie (1995), Scott (1995) and Mclagan (1995), that electronic dissociative recombination of the various $C_2H_3CO^+$ ions is unlikely to be an efficient way to form $H_2C_3O$ isomers because, if electronic dissociative recombination of the various $H_3C_3O^+$ isomers did preserve the carbon skeleton, the most stable $H_2C_3O$ isomer, propadienone ($H_2C=C=C=O$), should be abundant in molecular clouds.

The model results are in satisfactory agreement with observations for early cloud ages ($10^5$ years). In our model, propynal is mainly formed through the well-known $O + C_3H_3$ reaction, cyclopropenone is formed mainly through the $OH + c-C_3H_2$ reaction and propadienone is formed mainly through the $OH + l-C_3H_2$ reaction. This last reaction involves a small flux. The main destruction reactions are with carbon atoms and protonation as the DR reactions are not thought to reform $H_2C_3O$ isomers. It should also be noted that several reactions are very complex, particularly the $OH + c-C_3H_2$ reaction, and the determination of precise branching ratios requires a full theoretical study. The branching ratios in the DR reactions are critical as well. Although this is beyond the scope of the present paper, such calculations and/or experiments should be considered in the future.

JCL, KMH and VW thank the French CNRS/INSU program PCMI for their partial support of this work. VW researches are funded by the ERC Starting Grant (3DICE, grant agreement 336474). MA and JC acknowledge funding support from the European Research Council (ERC Grant 610256: NANOCOSMOS) and from Spanish MINECO through grants CSD2009-00038, AYA2009-07304, and AYA2012-32032.

We also thank the anonymous reviewer for his useful comments to improve the manuscript.

MOLPRO, version 2009.1, a package of ab initio programs, H.-J.Werner, P. J. Knowles, R. Lindh, F. R. Manby, M. Schütz, P. Celani, T. Korona, A. Mitrushenkov, G. Rauhut, T. B.